\documentclass[aps,
prl,
10pt,
twocolumn,
superscriptaddress,
longbibliography
]{revtex4-2}

\usepackage{xr}
\usepackage{amsmath,amssymb}
\usepackage{enumitem}
\usepackage{graphicx}
\graphicspath{ {Plots/} }
\usepackage{dcolumn}
\usepackage{bm}
\usepackage{wasysym}

\usepackage[dvipsnames]{xcolor}

\makeatletter
\renewcommand\frontmatter@abstractwidth{\dimexpr\textwidth-1.5in\relax}
\makeatother

\newcommand{\beq}{\begin{equation}}
\newcommand{\eeq}{\end{equation}}
\newcommand{\bea}{\begin{eqnarray}}
\newcommand{\eea}{\end{eqnarray}}

\renewcommand{\figurename}{Fig.}

\begin{document}

\title{ Symmetry Enforced Fermi Surface Degeneracies Observed in Time-Reversal Symmetry-Breaking Superconductor LaNiGa$_2$}

\author{Matthew Staab}
	\affiliation{Department of Physics and Astronomy, University of California, Davis, Davis, California $95616$, USA}
  \author{Robert Prater}
	\affiliation{Department of Physics and Astronomy, University of California, Davis, Davis, California $95616$, USA}
 \author{Sudheer Sreedhar}
	\affiliation{Department of Physics and Astronomy, University of California, Davis, Davis, California $95616$, USA}
  \author{Journey Byland}
	\affiliation{Department of Physics and Astronomy, University of California, Davis, Davis, California $95616$, USA}
 \author{Eliana Mann}
	\affiliation{Department of Physics and Astronomy, University of California, Davis, Davis, California $95616$, USA}
        \affiliation{John A. Paulson School of Engineering and Applied Sciences, Harvard University, Cambridge, MA, USA}
 \author{Davis Zackaria}
	\affiliation{Department of Physics and Astronomy, University of California, Davis, Davis, California $95616$, USA}
\author{Yunshu Shi}
	\affiliation{Department of Physics and Astronomy, University of California, Davis, Davis, California $95616$, USA}
\author{Henry J. Bowman}
        \affiliation{Department of Physics and Astronomy, University of California, Davis, Davis, California $95616$, USA}
	% \affiliation{Department of Physics and Astronomy, Carleton College, Northfield, MN 55057}
\author{Andrew L. Stephens}
        \affiliation{Department of Physics and Astronomy, University of California, Davis, Davis, California $95616$, USA}
 \author{Myung-Chul Jung}
	\affiliation{Department of Physics, Arizona State University, Tempe, AZ 85281, USA}
 \author{Antia S. Botana}
	\affiliation{Department of Physics, Arizona State University, Tempe, AZ 85281, USA}
 \author{Warren E. Pickett}
	\affiliation{Department of Physics and Astronomy, University of California, Davis, Davis, California $95616$, USA}
 \author{Valentin Taufour}
	\affiliation{Department of Physics and Astronomy, University of California, Davis, Davis, California $95616$, USA}
 \author{Inna Vishik}
        \email[Corresponding author: ]{ivishik@ucdavis.edu}
	\affiliation{Department of Physics and Astronomy, University of California, Davis, Davis, California $95616$, USA}

\begin{abstract}
LaNiGa$_2$ is superconductor that breaks time-reversal symmetry in the superconducting state without any known nearby magnetism. Recently, single crystals of LaNiGa$_2$ have been synthesized, revealing a nonsymmorphic Cmcm space group. Here, we report measurements of the electronic structure of LaNiGa$_2$ throughout the three-dimensional Brillouin zone (BZ) using angle-resolved photoemission spectroscopy (ARPES). Our findings show broad consistency with density functional theory (DFT) calculations and provide evidence for degeneracies in the electronic structure that are predicted from the space group. The calculations also predict four Fermi surfaces which cross the purported nodal plane and should therefore form two degenerate pairs. We report evidence for those predicted symmetry enforced degeneracies as well as accidental near degeneracies throughout the BZ. These degeneracies and near-degeneracies may play a role in the pairing mechanism of LaNiGa$_2$. Our results provide insight into the interplay between structure, Fermiology, and superconductivity in unconventional superconductors with nonsymmorphic space group.

\end{abstract}

\date{\today}

\maketitle

\section{Introduction}

Unconventional superconductors have mechanisms or phenomena that differ from superconductors described by Bardeen-Cooper-Schrieffer (BCS) theory.  The normal state electronic properties of unconventional superconductors can inform their superconducting mechanism as well as identify factors that raise or lower the transition temperature, T$_c$.  For example, the experimentally established Fermiology of iron based superconductors influenced early proposals of interband pairing~\cite{Kordyuk2012, Zabolotnyy2009}, and in cuprate high-temperature superconductors, the normal-state pseudogap presents one of the most formidable challenges to establishing the superconducting mechanism~\cite{Fradkin2012,Hashimoto2014}. Among superconductors that have been reported to break time reversal symmetry below T$_c$,  Sr$_2$RuO$_4$ has been the most heavily investigated, and its normal-state band structure, particularly the van hove singularity~\cite{Yokoya1996,Wang2013}, has been implicated in the origin of superconductivity and the evolution of T$_c$ under strain~\cite{Sunko2019}.  In other classes of superconductors, there are examples of isostructural materials that alternately break/maintain time reversal symmetry in the superconducting state, providing further indication that details of Fermiology can play a role in the pairing state~\cite{Shang2018}.

Time-reversal symmetry breaking (TRSB) superconductivity is relatively rare, with only several dozen single-material (not heterostructure) candidates identified to date~\cite{Ghosh2021,Badger2022}, and many of these compounds have magnetism or magnetic fluctuations which may play a role in mediating TRSB superconductivity~\cite{Romer2020,Aoki2022,Farhang2023}. LaNiGa$_2$ is a rare example where magnetism is absent due to the Ni bands being filled~\cite{Quan2022, Singh2012}.  This material has a superconducting transition temperature, T$_c\approx$2K~\cite{Zeng2002}, and most prior studies have been on polycrystalline specimens.  $\mu$SR experiments reported the onset of spontaneous magnetization below T$_c$, interpreted as demonstrating TRSB superconductivity ~\cite{Hillier2012}.

\begin{figure}%[hptb!]
	\centering
	\includegraphics[width=\linewidth]{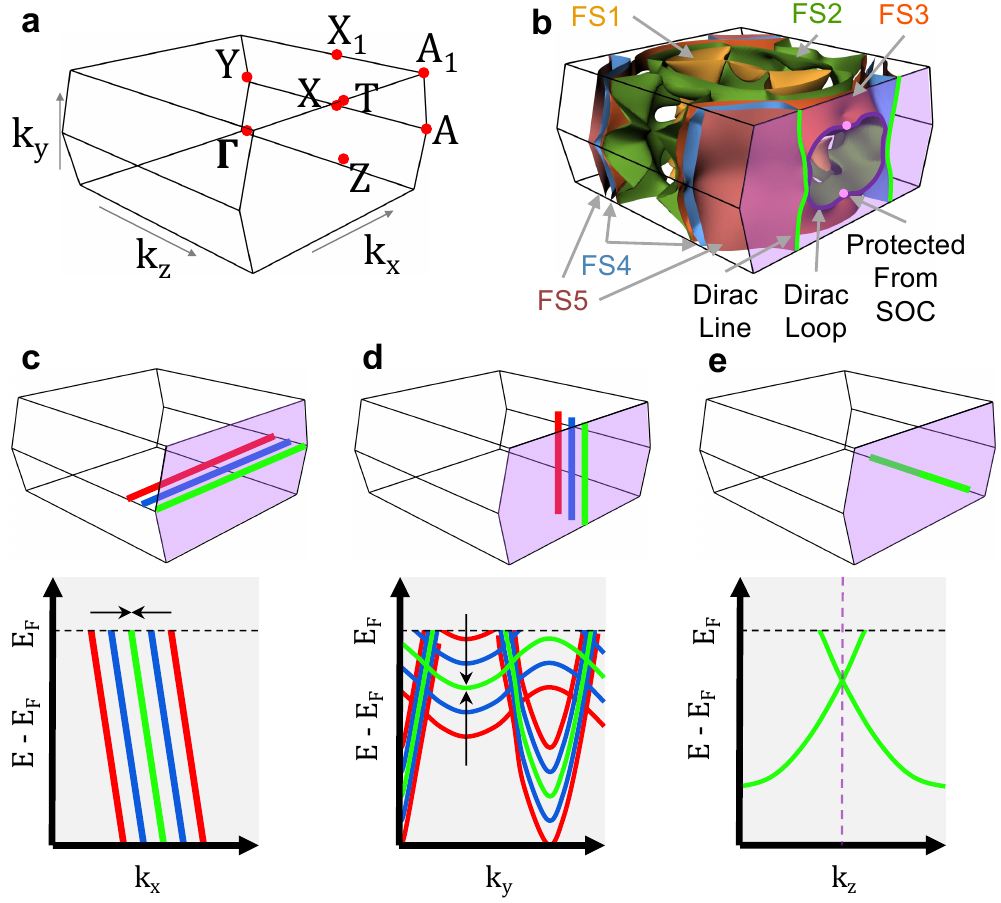}
	\caption{\textbf{Predicted Fermiology and Nodal Plane Features} (a) The C-Orthorhombic Brillouin zone with high symmetry points and axes labeled. The purported nodal plane is the $k_z$ boundary, or the $Z-T-A$ plane. (b) The five Fermi surfaces of LaNiGa$_2$ shown in the Brillouin zone, reproduced from Ref. \cite{Badger2022}. The pink dots indicate the SOC protected nodal points.  Features labeled `nodal line' and `nodal loop' are degenerate crossings at $E_F$ that get split by SOC. (c)-(e) Schematic of predicted band structure degeneracies. (c)-(d) cuts parallel to the nodal plane are expected to show band merging in energy and momentum. (e) When piercing the predicted nodal plane, parallel to $k_z$, the calculations predict a Dirac crossing of bands. In (b-e) The purple plane is the purported nodal plane.
 }
	\label{fig:TheorySchematic}
 \end{figure}

The recent synthesis of single crystals has enabled a wider array of experiments to understand structural, thermodynamic, and electronic properties relevant to the superconducting mechanism~\cite{Badger2022}. More than enabling new experiments, these single crystals demonstrated that the space group was previously mis-identified as Cmmm, with the correct space group being Cmcm, which is nonsymmorphic. The Cmcm space group imposes symmetry-enforced degeneracies at the $k_z$ Brillouin zone boundary ($Z-T-A$ plane, Fig.~\ref{fig:TheorySchematic}(a)-(b)). Spin-orbit coupling (SOC) splits this degeneracy everywhere except for the $k_x=0$ line on that plane, but only slightly~\cite{Quan2022}. The symmetry enforced degeneracies and near-degeneracies of this space group are predicted to contribute to the pairing mechanism in LaNiGa$_2$~\cite{Quan2022}, and may be an important for other inversion-symmetry-preserving superconductors with nonsymmorphic space groups.

Other materials in the Cmcm space group have recently garnered attention due to the degenerate features of their band structures. TaNiTe$_5$ and TaPtTe$_5$ are two such examples of non-superconducting Cmcm crystals that have been shown by ARPES to have Dirac nodal lines at the BZ boundary~\cite{Hao2021, Xiao2022}. Being non-superconducting, these compounds show that it is not just the space group and related Fermi surface degeneracies that make LaNiGa$_2$ a superconductor.

Here we present the experimental normal-state electronic structure of TRSB superconductor LaNiGa$_2$ throughout the entire 3D BZ, as measured by ARPES.  The following results are presented: (1) broad agreement between density functional theory (DFT) and measured electronic structure throughout 3D BZ. (2) Evidence for Fermi surface degeneracies on $Z-T-A$ plane, within resolution of experiment, via distinct bands merging into one approaching this plane in parallel (Fig.~\ref{fig:TheorySchematic}(c)-(d)). (3) Evidence of predicted linear-crossings of all bands when cutting perpendicular to $Z-T-A$ plane (Fig.~\ref{fig:TheorySchematic}(e)). (4) Evidence for near-degeneracies of bands throughout the BZ separate from those on the $Z-T-A$ plane.  The relevance of these features to the proposed mechanism of unconventional superconductivity will be discussed.

\section{Methods}

{\textbf{Crystal growth and ARPES Measurements}} 
 The samples were grown using the technique described in the original single-crystal growth by Badger et al.~\cite{Badger2022}.
 
ARPES measurements were performed at Stanford Synchrotron Radiation Lightsource (SSRL) beamline $5-2$.  Samples were cleaved \textit{in-situ} at pressure better than $5 \times 10^{-11}$ Torr and at temperatures less than $15K$. The photon energy was varied as stated in relevant figures.  The energy and momentum resolution of the experiments were 25 meV  and $ \approx 0.01\AA ^{-1}$ respectively. The beam spot size was approximately $ 20 \mu m \times 30 \mu m$ ('microARPES') for all experiments.

\begin{figure}%[hptb!]
	\centering
	\includegraphics[width=3in]{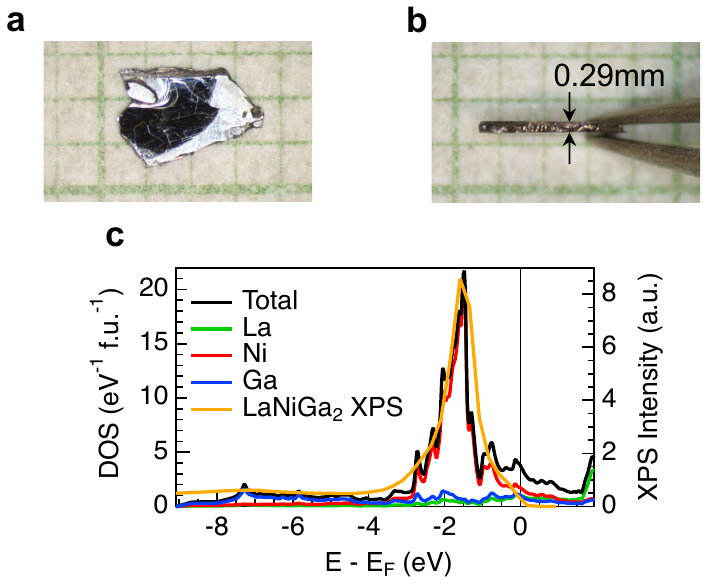}
	\caption{\textbf{Sample Dimensions and XPS} 
 (a) LaNiGa$_2$ sample pictured on millimeter paper showing typical shape and size of the platelet crystals in the a-c plane (natural orientation). (b) Image of the sample held on its side (thin-edge orientation). 
 (c) Projected density of states calculations along with XPS measured with $144eV$ photons in the natural orientation.
 }
	\label{fig:XtalPicsXPS}
\end{figure}

LaNiGa$_2$ single crystals grow in a platelet shape with typical dimensions of $ 2mm \times 2mm$ (figure~\ref{fig:XtalPicsXPS} (a)-(b)).
The platelet-shaped samples are most naturally cleaved $in\text{-}situ$ by mounting a ceramic top post on top of the platelet, hereafter called the `natural cleave' orientation.  Cleaving the crystal in the natural orientation allows for ARPES that approximately probes of planes of constant $k_y$. 
% Both the sample-to-substrate as well as the sample-to-top-post adhesives were silver epoxy. 
In order to probe band structure at roughly constant values of $k_z$ or $k_x$ without relying exclusively on the lower-resolution photon energy tuning method, we mounted the platelet samples on their thin-edge. This was done by etching a groove into the sample holder and adhering the sample into the groove. The samples were mounted into the groove with ultrahigh vacuum compatible Locktite, and a conductive pathway created with the silver epoxy Epo-tek H21D. 
% We developed a method (\emph{thin-edge}) to mount the platelet samples along their thin-edge by etching a grove into the mounting plate and setting the sample on its edge within the groove. 
For data shown here, the samples were pre-aligned using Laue~\cite{SI} to have the $\vec{a}$ axis normal to the sample holder.  The thin-edge-mounted samples were broken $in\text{-}situ$ using a flat-edged wobble stick to apply uniform pressure across the face without using a top post. This configuration of thin-edge cleaving was used to probe the $k_y$ dependence of the electronic structure directly with ARPES. MicroARPES is crucial for enabling measurements of thin-edge-cleaved samples.

{\textbf{Electronic Structure Calculations}} 

Density functional theory calculations were carried out using the full-potential linearized augmented planewave code {\sc{Wien2k}}~\cite{Blaha2020wien2k}. We used the Perdew-Burke-Ernzerhof (PBE) version of the generalized gradient approximation as the exchange-correlation functional~\cite{Perdew1996generalized}. The atomic coordinates used for these calculations were those listed in~\cite{Badger2022, Quan2022}. The muffin-tin sphere sizes were  2.50 a.u for La; 2.32 a.u for Ni, and 2.20 a.u for  Ga. The plane wave cutoff was determined by $R_{mt}^{min}K_{max}=7$, and the $k$-point mesh used was $14\times 14\times 14$ for self-consistency and $34\times 34\times 33$ for the projected density of states. The effects of spin-orbit coupling were included by using the second variational method as implemented in {\sc{Wien2k}}.

\section{Results}
\begin{figure}%[hptb!]
	\centering
	\includegraphics[width=\linewidth]{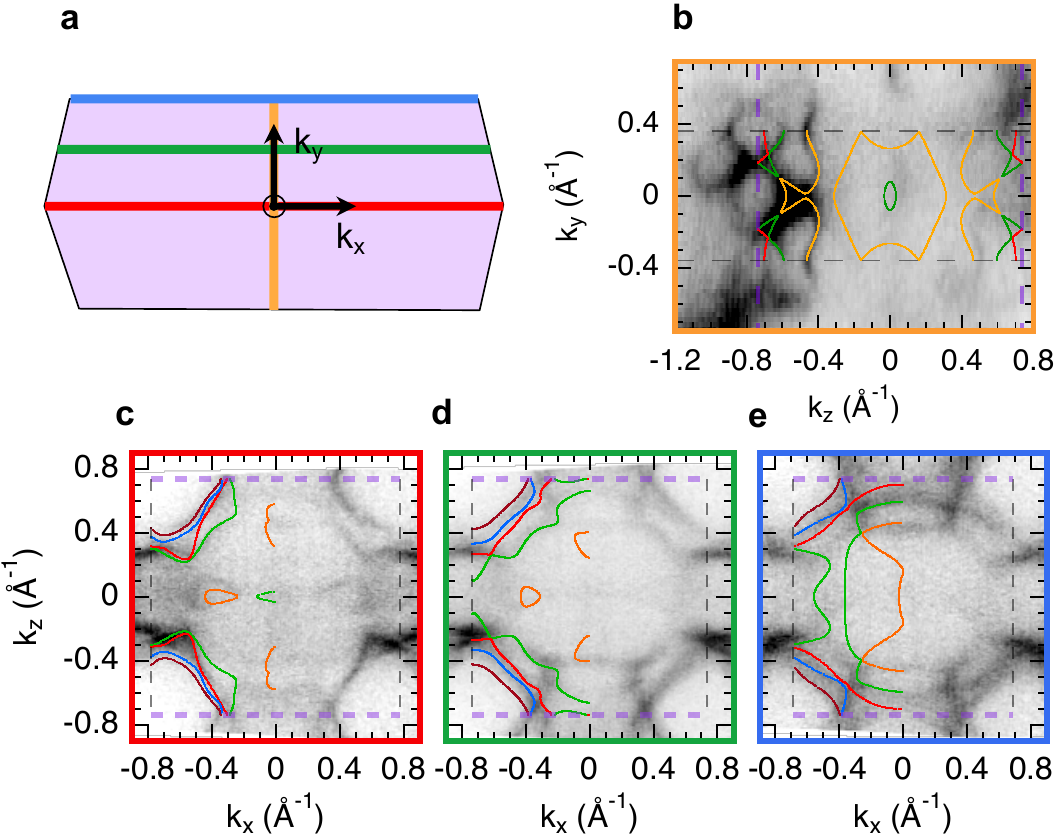}
	\caption{\textbf{Observed Fermiology for selected planes in BZ.} (a) Projection of the BZ showing locations of the planes with colored lines corresponding to the borders in (b-e). (b) Constant energy map integrated $45meV$ around the Fermi energy at $144eV$ in the thin-edge orientation. (c-e) ARPES constant energy maps integrated $50meV$ around the Fermi energy. Measured in the natural platelet orientation at $116eV$, $124eV$ and $132eV$ respectively to access the center and top of the BZ. DFT calculated Fermi crossings shown in (b-e) are colored corresponding to the Fermi surfaces in Fig.~\ref{fig:TheorySchematic}. In (c-e) they are only overlaid on half of the zone for clarity of the data. Purple plane in (a) and dashed lines in (b-e) indicate $Z-T-A$ plane location. Black dashed lines indicate the other BZ.}
	\label{fig:Fermiology}
\end{figure}

Fig.~\ref{fig:XtalPicsXPS}(c) shows an angle-integrated valence band spectrum, together with calculated partial and total density of states.  These x-ray photoelectron spectroscopy (XPS) data are collected on the same natural cleave as ARPES data.
Maps of constant energy at the Fermi level ($E_F$), yielding Fermi surface maps, are shown  for four characteristic planes in momentum space in Fig.~\ref{fig:Fermiology}. The associated DFT calculations are partially superimposed to establish the alignment between theoretical and experimental Fermiology. The displayed planes of the BZ are schematically shown in panel (a) by orange, red, green, and blue lines which correspond to (b) $k_x = 0$ (144 eV), (c) $k_y = 0$ (116 eV)), (c) $k_y = 0.516\cdot 2\pi/c$ (124 eV), (e) $k_y = 2\pi / c$ (132 eV) respectively.  Data for panel (b) were collected in the thin-edge orientation directly probing the $k_x = 0$ plane, which permits spectra orthogonal to the purported nodal plane. Broader photon energy dependence was preformed from $110eV$ to $144eV$, and is shown in the the SI, where methods for establishing the relationship between photon energy and $k_{\perp}$ (momentum perpendicular to surface plane) is also discussed~\cite{SI}.

\begin{figure}%[hptb!]
	\centering
	\includegraphics[width=\linewidth
 ]{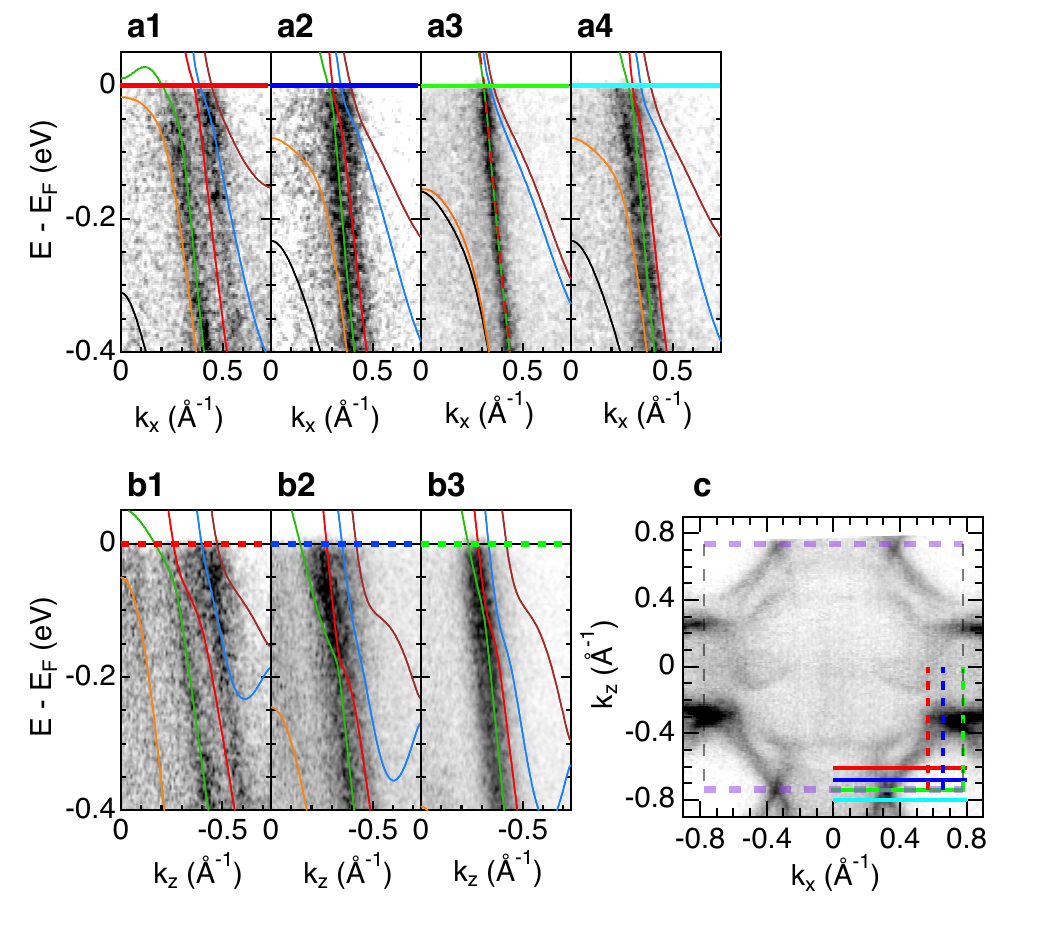}
	\caption{\textbf{Band merging in momentum at zone boundaries.} (a1-a4) Spectra approaching (a1,a2) at (a3) and beyond (a4) the $Z-T-A$ plane as a function of $k_z$. (b1-b3) Spectra approaching (b1,b2) and on (b3) the BZ boundary along $k_x$. (c) Constant energy map at the Fermi level with $120eV$ photons indicating the location of the spectra in (a) and (b). (a-b) DFT predicted bands are overlaid according to the color of the FS they create in Fig.~\ref{fig:TheorySchematic}.}
	\label{fig:NodalAndOrthogonal}
\end{figure}

As discussed, one defining feature of the predicted electronic structure is pairs of bands merging together when approaching the $Z-T-A$ plane. Fig.~\ref{fig:NodalAndOrthogonal} shows that merging as a function of momentum both approaching the purported $Z-T-A$ plane as well as the orthogonal zone boundary at a photon energy of $120eV$.

Panels (a1-a4) show the spectra approaching (a1,a2), on (a3) and past (a4) the predicted nodal plane. Panel (b) similarly shows the spectra approaching (b1, b2) and on (b3) the zone boundary orthogonal to the predicted nodal plane. Panel (c) shows the described locations of the cuts in panel (a) and (b) on-top of a constant energy map integrated $\pm50meV$ around the Fermi level.

Panel (a) has two separated bands in panels (a1, a2) that appear to merge in panel (a3) and split again in (a4).

 Panel (b) also shows two well separated bands in panel (b1) that closely approach each other in panel (b3). Whereas spectra in (a2) and (b2) have similar linewidth, the spectrum in (a3) is considerably sharper than (b3). Calculated electronic structure for these cuts is overlaid with their color corresponding to the Fermi surface number given in Fig.~\ref{fig:TheorySchematic}.
 
\begin{figure}%[hptb!]
	\centering
	\includegraphics[width=\linewidth]{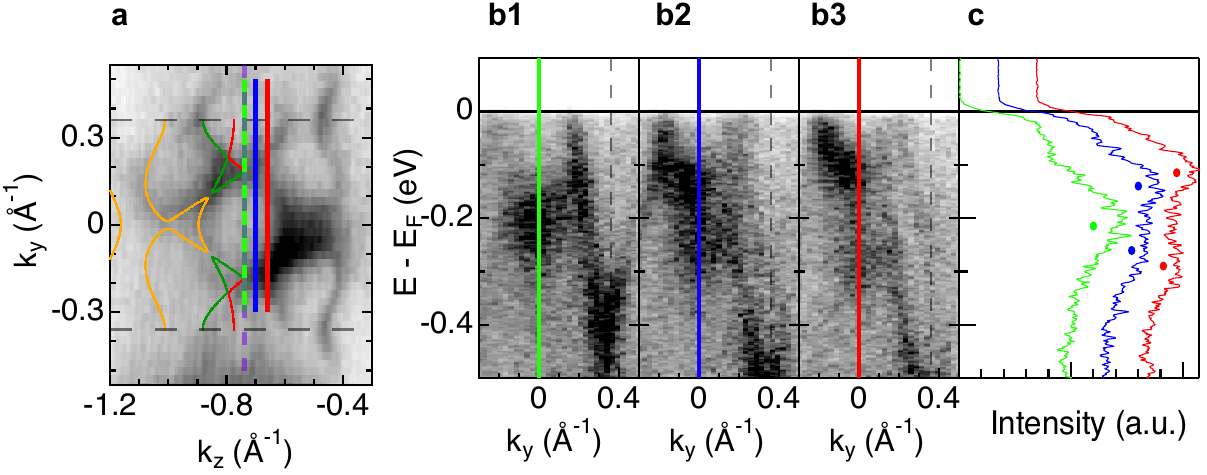}
	\caption{\textbf{Band merging in energy, thin-edge cleave.} (a) Fermi level constant energy contour cropped from Fig.~\ref{fig:Fermiology} (b) showing locations of cuts in (b1)-(b3). Black and purple dashed lines indicate BZ boundaries. (b1)-(b3) ARPES spectra at three parallel cuts (c) $k_y=0$ EDCs of panels (b1)-(b3) in green, blue, offset for clarity.  Dots are guide-to-the-eye for peak position Thin Edge: LA\_0151 From \date{May 11, 2022}``20220511\_LaNiGa2"}
	\label{fig:Nodal_Plane_Thin}
\end{figure}

With the thin-edge cleave perpendicular to $\vec{a}$, we can visualize bands merging in energy approaching the $Z-T-A$ plane, because the observed bands have band bottoms at lower binding energy. In Fig. ~\ref{fig:NodalAndOrthogonal} we only discussed merging in momentum because the band bottoms are well below the displayed energy range.  Nevertheless, for cuts approaching the purported nodal plane in parallel, bands always merge in both energy and momentum. Figure~\ref{fig:Nodal_Plane_Thin} shows spectra along three parallel cuts moving progressively further from the zone boundary (panels (b1-b3)).  In panel (c), we compare energy distribution curves (EDCs) at the $k_y=0$ point, where a band bottom is located.  Panel (b1), taken along the zone boundary, shows a single band in the spectrum and a single peak in the EDC. As one moves away from the zone boundary, the bands split in two as seen both in the image plots and the EDCs. The magnitude of this split is $\approx200 meV$ for cuts differing by $\approx6\%$ of the BZ.

\begin{figure}%[hptb!]
	\centering
	\includegraphics[width=\linewidth]{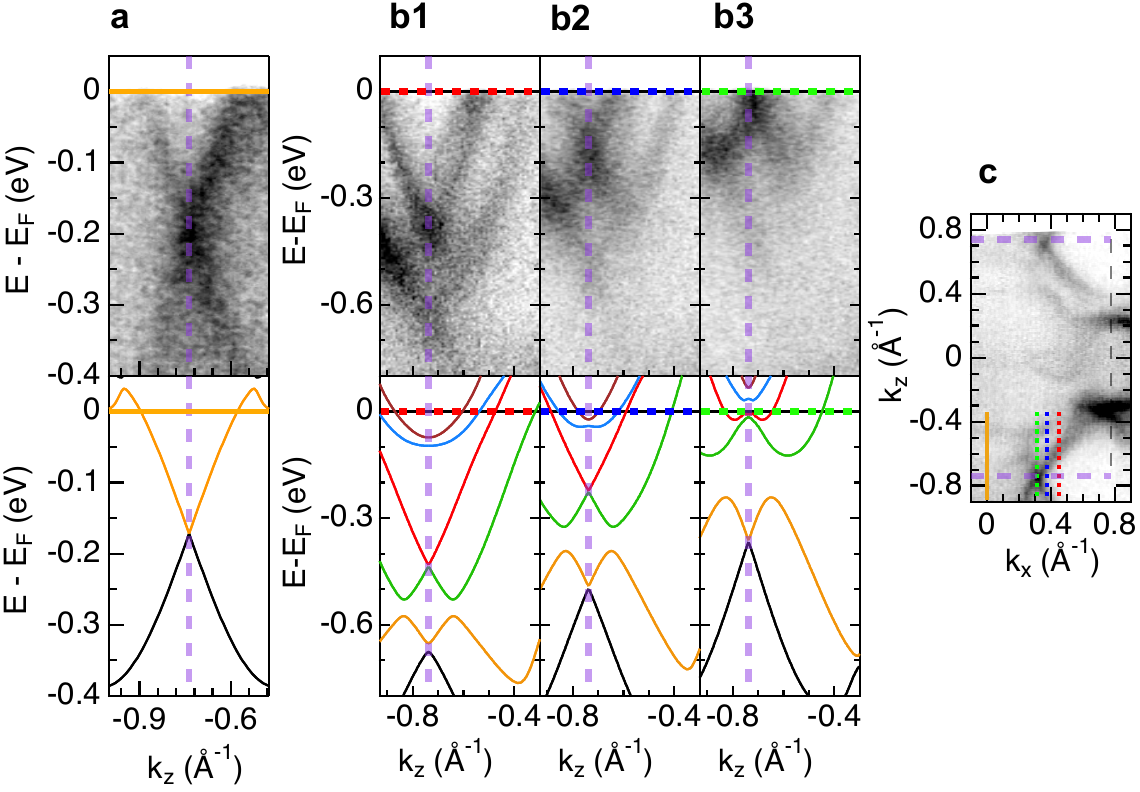}
	\caption{\textbf{Linear crossings for cuts perpendicular to nodal plane.} (a) Thin edge orientation spectrum taken along the $\Gamma - Z$ line at a photon energy of $144eV$. (b) Spectra taken in the natural orientation with $120eV$ photons near a Fermi crossing at the zone boundary. (c) Constant energy contour from the natural orientation indicating the location of cuts in (a-b). The cut from (a) is along the $k_y = 0\AA^{-1}$ line while (b-c) are from the $k_y \approx 0.09\AA^{-1}$ plane. (a-b)(bottom) DFT predicted bands are shown according to the color of the FS they create in Fig.~\ref{fig:TheorySchematic}.}
	\label{fig:DiracCrossings}
\end{figure}

So far we have provided evidence for band degeneracy at the $Z-T-A$ plane by investigating the spectra along lines parallel to the $Z - T - A$ plane. The second predicted signature of the $k_z$ BZ boundary is that for all cuts perpendicular to it, all bands which cross the $Z-T-A$ plane should do so with a Dirac-like (linear) crossing, up to SOC. In Fig.~\ref{fig:DiracCrossings} (a) we show a spectrum along the $k_z$ axis that shows a Dirac-like crossing at the $Z-T-A$ (purple dashed line) $0.2eV$ below $E_F$. That spectrum, being along the $k_z$ axis, intercepts the nodal plane along the $Z-T$ line and is therefore predicted to be protected against SOC splitting due to an additional symmetry. In panels (b1) - (b3) of Fig.~\ref{fig:DiracCrossings} we show three parallel cuts that intersect the $Z-T-A$ plane successively closer to the purported nodal line.  In the experimental spectra, two linear crossings are clearly seen in (b1), which successively move to lower binding energy approaching the momentum marked by the `X'-shaped feature in the Fermi surface map (c). Finally, at the purported nodal line (b3), a linear crossing is observed right at $E_F$. The crossings in (b) are not protected from SOC and are therefore gapped, but their gap size too small to be resolved in this work. The bottom of panels (b)-(c) show the DFT calculated dispersion in the corresponding locations.

\section{Discussion}

We begin by addressing the lack of magnetism in LaNiGa$_2$, previously justified by filled Ni-3\textit{d} bands. In DFT calculations of density of states (DOS), these filled bands yield a large peak at $\approx 1.8 eV$ binding energy in Fig.~\ref{fig:XtalPicsXPS} (c). The measured XPS intensity, a rough proxy for DOS,  is consistent with this calculated feature, supporting the stated absence of magnetism in this compound. We also note that recent nuclear magnetic/quadrupolar resonance and magnetization measurements in the normal state further do not find evidence of magnetic fluctuations or enhanced paramagnetism \cite{sherpa2023absence}.

We now turn to the measured band structure, where we observe overall good agreement between DFT and experiment throughout the 3D BZ. We note strong matrix element effects~\cite{MOSER201729}, which is the reason why certain bands are observed more, less, or not at all in certain regions of the BZ in the present data. There do exist a few discrepancies between measured and predicted bands, particularly in the most 3-dimensional bands which form FS 1 and 2. This can be seen in Fig.~\ref{fig:Fermiology} (c) where the `teardrop' shaped pocket of FS 1 near $k_x = -0.4\AA^{-1}$ is observed to be larger than predicted. Further, in panel (d) we observe elongated features of FS 1 and 2 that appear similar to those of panel (e).
The more 2D Fermi surfaces, FS 3 and 5, show no observed disagreement with the predicted electronic structure. Additionally, FS 1 and 2 show strong agreement with predictions in the thin-edge cleave shown in Fig.~\ref{fig:Fermiology} (b). 

The disagreements seen of FS 1 and 2 are then either: (1) real features of the electronic structure not captured by the calculations, and constrained to only small portions of the BZ or (2) artifacts arising from the photoemission process due to momentum uncertainty in the $k_\perp$ direction. Arguing against scenario (1), there is very good agreement between ARPES Fermi surfaces and DFT with regards to FS1 and 2 in the thin-edge cleave data.  Elaborating on (2), the short inelastic mean free path (IMFP) of electrons makes ARPES a surface sensitive measurement in the present range of photon energies. We calculate the IMFP of electrons with kinetic energy of $110eV$ traveling through LaNiGa$_2$ to be $\lambda = 5 \AA$ using a common empirical method method~\cite{TANUMA1987L849}. This IMFP leads to an appreciable quantum uncertainty of the perpendicular momentum $\Delta k_\perp = 1/\lambda = 0.2\AA^{-1}$~\cite{STROCOV200365}, comprising $\approx 25\%$ of the BZ along $k_y$. 
Additionally, in ARPES measurements there can be preferential sampling of bands which are being probed by the spectrometer slit normal to their constant-energy surface. These two effects: (1) $k_\perp$ broadening and (2) extremum-selective sampling may explain the discrepancies observed in Fig.~\ref{fig:Fermiology} as the extremum position observed for FS 1 is at $k_y \approx 0.016\AA^{-1}$.

Now we turn to focus on the observed features at the purported nodal plane.  The first observation is bands merging when approaching the $Z-T-A$ plane with parallel cuts (Fig. \ref{fig:NodalAndOrthogonal},\ref{fig:Nodal_Plane_Thin}).  While the calculations predict SOC splitting of the bands along the nodal plane at all positions except the $Z-T$ line, the splitting is extremely small in most locations. In Fig.~\ref{fig:NodalAndOrthogonal} (a3) the splitting between FS 2 and 3 is $5meV$, much smaller than our experimental resolution. Also predicted are Dirac-like crossings for all bands when piercing the nodal plane in perpendicular.  This is observed for several distinct bands in Fig. \ref{fig:DiracCrossings} Momentum-space band merging can be additionally visualized in Fermi surface maps, particularly in Fig.~\ref{fig:Fermiology} (c)-(e), Fig. \ref{fig:NodalAndOrthogonal}(c), and many other examples shown in the SI \cite{SI}. These show an  `X'-shaped feature at the BZ boundary along one direction ($k_z$) but not along the other ($k_y$), consistent with the location of the nodal plane. Notably, most of these observations are discernible from data alone, independent from comparison to DFT calculations.

Beyond the nodal plane, we also present evidence for near-degeneracies throughout the BZ. These near-degeneracies are particularly persistent at the orthogonal zone boundary ($k_x$). FS 2 and 3 have near degeneracies not only along the orthogonal BZ boundary but within the BZ at many values of $k_y$. Near-degeneracies are observed in both the thin-edge Fermi surface as well as broad ranges of the photon energy dependent natural-cleave Fermiology. 
Four clear examples can be seen of near-degeneracies within the BZ in Fig.~\ref{fig:Fermiology}: First, panel (b) shows degeneracy of FS 2 and 3 at $k_z \approx -0.65\AA^{-1}$. Second, panel (c) shows a near-degeneracy of FS 2 and 3 at $k_x \approx -0.55\AA^{-1}$ where the bands are almost stuck until they reach the $k_x$ zone boundary. Third, panel (d) shows FS 2 and 3 again nearly degenerate and then stuck near the $k_x$ boundary. Finally, panel (e) again shows FS 2 and 3 being nearly degenerate inside the BZ at $k_z \approx -0.35\AA^{-1}$ nearly co-located with a near degeneracy of FS 3 and 4.

We now connect these results to prior work on this material.  Early experiments have laid out a set of constraints for the pairing symmetry and mechanism of LaNiGa$_2$.  The point group symmetry of the crystal structure, which remains unchanged with the newly assigned space group, only permits four order parameters that break time reversal symmetry~\cite{Hillier2012,Quan2022}.  All are non-unitary and all have weak SOC, the latter constraint which is confirmed with the present data.  Additionally, heat capacity and superfluid density were shown to be inconsistent with nodes of any kind~\cite{Weng2016,Badger2022}.  The latter showed better consistency with a two-gap gamma model than a one-gap BCS fitting, which was more recently bolstered by transverse-field $\mu$-SR measurements \cite{sundar2023gap}. 

Together, these results point to a fully-gapped triplet state.  The superconducting wavefunction must be odd overall to comport with Fermion exchange, so an even spin component implies an odd spatial component.  This presents a problem when one includes information about the Fermiology, as an odd spatial wavefunction must change sign across the $\Gamma$ point, so a Fermi surface that encloses $\Gamma$ should have a node, inconsistent with experiments.  The calculated Fermiology based on both the previous Cmmm~\cite{Singh2012} and refined Cmcm~\cite{Badger2022,Quan2022} crystal structure includes Fermi surfaces which envelop the gamma point.  In the present calculation, this includes a small pocket of FS1, as well as FS2 and 3.   Previously, this apparent inconsistency was resolved by considering the multiband nature of this material, wherein triplet pairing between nearly degenerate Fermi surfaces can ensure an overall odd wavefunction~\cite{Ghosh2020}, if the order parameter changes sign upon swapping band index.  This proposed internally antisymmetric nonunitary triplet pairing (INT) state appears to be consistent with all experimental results about the superconductivity in LaNiGa$_2$ to date: TRSB below T$_c$ \cite{Hillier2012}, nodeless superconducting gap \cite{Weng2016,Badger2022}, consistency with two gaps \cite{sundar2023gap}, and insensitivity of superconductivity to non-magnetic impurities \cite{ghimire2023electron}.

The present work adds experimentally-motivated plausibility to that argument, and proposes further experiments to understand and optimize superconductivity in LaNiGa$_2$.  The Cmcm space group imposes symmetry-enforced degeneracies along one plane of the BZ, the so-called ‘nodal plane.’ Our experiments find support for this in three ways: bands merging in momentum approaching the nodal plane with parallel cuts (Fig.~\ref{fig:NodalAndOrthogonal}), bands merging in energy approaching the nodal plane with parallel cuts (Fig.~\ref{fig:Nodal_Plane_Thin}), and all cuts perpendicular to the nodal plane yielding Dirac-like crossings for all bands at all energies (Fig.~\ref{fig:DiracCrossings}).  Although SOC lifts this degeneracy on most of the nodal plane, there are still three sources of degeneracy and near-degeneracy to support interband pairing: (1) symmetry-enforced exact degeneracy at predicted nodal points along $Z-T$ line (2) near-degeneracy on nodal lines and nodal loop arising from symmetry coupled with a small (~meV) amount of SOC (3) accidental near-degeneracies including on the $k_x$ BZ boundary (Fig.~\ref{fig:NodalAndOrthogonal}(b)).  The latter were present in band structure calculations based on the previous Cmmm structure, but the former two are a consequence of the Cmcm structure.  The overall strong agreement between DFT calculations and experiment throughout the 3D BZ further supports the prediction of a nodal plane, as well as the presence of only weak SOC~\cite{Hillier2012}, which is required for the experimentally-compatible pairing states.

Despite the growing body of knowledge on LaNiGa$_2$, accelerated by the availability of single crystals, questions still remain.  Interband pairing benefits from near-degeneracies throughout the BZ; the ones on the nodal plane will remain as long as the crystal structure is unchanged, but the ones outside the nodal plane may be tunable by uniaxial strain.  The effect of this tuning on T$_c$ can further confirm the pairing mechanism that appears to be supported by a plurality of experiments.  Moreover, it can clarify if there are pairing interactions outside of electron-phonon~\cite{Tutuncu2014} which is the primary candidate in LaNiGa$_2$, such as originating magnetic phases which often accompany TRSB superconductor. As an example of this, LaNiC$_2$ was also thought to be entirely non-magnetic, until pressure-dependent studies on single-crystals revealed a hidden antiferromagnetic regime~\cite{Landaeta2017}.  

Another important direction is substitutional carrier tuning, which is expected to maintain the nodal points, maintain near-degeneracies on the nodal plane, but tune the presence of small hole-like pockets throughout the BZ, clarifying their role in superconductivity, which is heretofore not considered in detail. 
These new experiments should be pursued in conjunction with revisiting foundational results about this material.  In other materials, claims of TRSB superconductivity have become less clear as more experimental evidence emerged, most prominently in Sr$_2$RuO$_4$ \cite{Pustogow2019,Chronister2021SRO}. With that insight, we emphasize the importance of repeating zero field $\mu$SR measurements, as well as other experiments sensitive to TRSB in the superconducting state, on single crystals of LaNiGa$_2$ where the surface/volume ratio is smaller and the constituency of impurities and inclusions may also be different.

Unconventional superconductors are a very broad class likely to contain a diverse set of novel pairing mechanisms. The subset of those that break time reversal symmetry is quite small, and the subset of these TRSB superconductors that do not contain nearby magnetism is smaller yet. The present high quality data, including in the unconventional thin-edge cleave geometry demonstrates that LaNiGa$_2$ can continue to yield experimental insights on momentum space electronic structure and can serve as a benchmark for this set of rare superconductors. One other example of TRSB superconductivity without nearby magnetism is $4$Hb-TaS$_2$. However, being a 2D Van der Waals material, $4$Hb-TaS$_2$ likely does not allow for a thin-edge cleave to be achieved in the same way that LaNiGa$_2$ is shown to here. 

\section{Conclusions}
We have performed the first comprehensive ARPES study of purported TRSB superconductor LaNiGa$_2$ throughout the 3D BZ.  This material yields high quality ARPES spectra. Additionally, we have demonstrated thin-edge cleaving in platelet-like specimens to yield ARPES spectra of comparable quality to conventional cleaving, with implications for further surface spectroscopy studies of platelet specimens.  Superconductivity is a low-energy phenomena which necessarily is affected by the Fermiology of a material. This is particularly true in the case of LaNiGa$_2$ which has an expected gap size of $k_B T_c \approx 0.6meV$~\cite{Quan2022}. The Cmcm space group enforces band degeneracies at the $k_z$ boundary, which constrains any Fermi surface that crosses that plane to be degenerate with another. LaNiGa$_2$ has 4 Fermi surfaces which cross the $Z-T$ plane, two of which cross the additionally protected $Z-T$ line and are degenerate even with SOC. These degenerate Fermi surfaces are predicted to provide the platform which allows for the interband superconductivity.

Our results have given evidence that predicted degenerate Fermi crossings are present within our measurement resolution, and also shown that in other regions of the BZ there are symmetry enforced as well as accidental nearly-degenerate crossings. These nearly degenerate crossings are most prominent at the $k_x$ boundaries.  Known superconductors in nonsymmorphic space groups are a fruitful avenue to search for TRSB~\cite{Badger2022}, and LaNiGa$_2$ is emblematic of the important interplay between normal-state Fermiology and superconductivity inherent in this approach.

{\textbf{Data Availability}} The ARPES data used in this manuscript will be available following publication.

{\textbf{Acknowledgments}}
We thank Filip Ronning, Donghui Lu, Makoto Hashimoto, and Peter Abbamonte for helpful discussions.  The synthesis and characterizations were supported by the UC Laboratory Fees Research Program (LFR-20-653926). The natural-cleave measurements in this manuscript was supported by AFOSR Grant No. FA9550-18-1-0156, while the thin-edge cleave measurements were supported by the Alfred P.Sloan Foundation(FG-2019-12170). Use of the Stanford Synchrotron Radiation Lightsource, SLAC National Accelerator Laboratory, is supported by the U.S. Department of Energy, Office of Science, Office of Basic Energy Sciences under Contract No. DE-AC02-76SF00515. H.B. was supported by the NSF-REU program PHY2150515. ASB was supported by the Alfred P. Sloan Foundation (FG-2022-19086). MS and SS acknowledge partial support by the advanced lightsource (ALS) graduate fellowship.

\bibliographystyle{apsrev4-2} 
\bibliography{main}

\onecolumngrid
\pagebreak
\renewcommand{\figurename}{Fig.}
\renewcommand{\thefigure}{S\arabic{figure}}
\setcounter{figure}{0}

\section{Supplemental Information}

\section{Reconstructed vs Direct energy vs momentum spectra}

In this work, much of the data were collected as Fermi surface maps in deflector mode, from which arbitrary energy vs momentum cuts can be constructred.  Two types of energy vs momentum spectra are used in the manuscript.  Direct spectra have their parallel momentum (x-axis) along the detector slit.  In contrast, reconstructed spectra are produced by stacking EDCs from varying deflector angles together, in order to produce cuts in directions perpendicular to direct ones. The reconstructed spectra in the manuscript are the following: Fig.  4 (b) panels, Fig.  5 (b) panels, Fig.  6 (b) panels. A similar reconstruction procedure can be used to reproduce a plane in the BZ perpendicular to the cleavage plane, by stacking together spectra collected with different photon energy.

\section{Crystal Alignment}

\begin{figure*}[h!]
	\centering
	\includegraphics[width=4in]{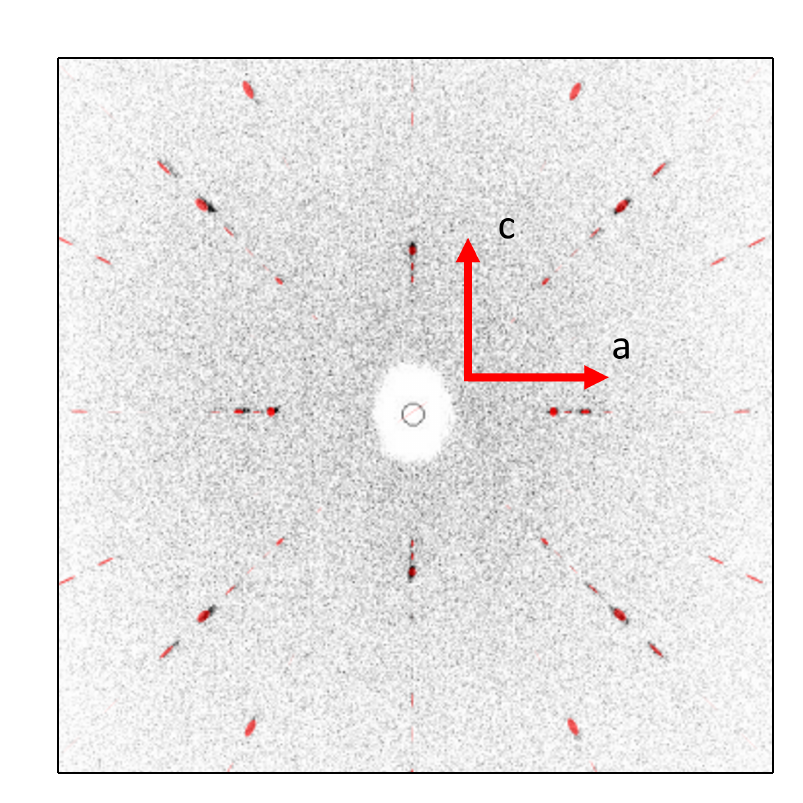}
	\caption{Laue image of LaNiGa$_2$ sample showing the orientations of the a and c axes. The red dots are the simulated diffraction peaks, taking into account both the X-ray source intensity and the structure factor. The simulation well distinguishes between the a and c axes. This orientation allows for samples to be mounted on the thin edge with the desired axis normal to the sample holder.}
	\label{fig:Laue}
\end{figure*}

Laue diffraction utilizes reflections of a polychromatic X-ray beam and those X-rays whose wavelength satisfy the Laue equations (Bragg condition) with the lattice are coherently reflected and form bright spots on the detector. These diffraction spots can then be used to easily orient the crystal, especially if one knows the crystal symmetry beforehand. In Fig.  \ref{fig:Laue}, we show the Laue image for LaNiGa$_2$. After alignment, samples are scored along the desired axis in order to preserve alignment when performing ARPES.  This is especially critical for mounting samples in a known orientation for the thin-edge cleave. 

\section{Determining $k_\perp$}
ARPES uses conservation of energy as well as momentum to probe electronic structure of materials. In the case of momenta parallel to the surface, there is no change in potential along that direction so momentum conservation is trivial, modulo a momentum kick from the photon which is only relevant at much higher photon energy than presently employed. However, in the case of momentum perpendicular to the surface, there is a potential energy discontinuity due to the termination of the crystal. That additional change provides an extra term to the relevant energy for determining the momentum. As given in equation \ref{eqn:kperp} below the additional term $V_0$, called the inner potential.  This can be determined experimentally by observing the periodicity of electronic structure as a function of photon energy \cite{Damascelli2004}. However, as discussed in the manuscript, the extreme surface sensitivity of the present experiments tends to mute dispersions in the $k_{\perp}$ direction, so additional methods are used.

\begin{equation}
\label{eqn:kperp}
    k_\perp = \frac{1}{\hbar} \sqrt{2m(E_{kin} \cos^2\theta + V_0})
\end{equation}

\begin{figure*}[h!]
	\centering
	\includegraphics[width=5in]{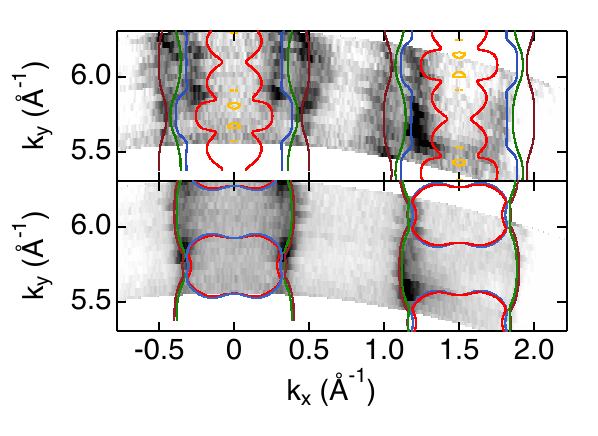}
	\caption{\textbf{Photon energy dependence of and near the $Z-T-A$ plane.} The top and bottom panel show constant binding energy contour ARPES maps over a range of photon energies from $110eV$ to $144eV$. (Top) $k_x$ near the BZB boundary, at the location shown by the solid red line in main text Fig.  4 (c). (Bottom) On the $k_z = \pi/c$ plane, probing the SOC split Dirac lines and loop. Overlaid on both are the DFT calculated Fermi crossings.}
	\label{fig:kz_nodal_hv_dep}
\end{figure*}

We determine the value of the inner potential to be $15eV$in order to align features seen in the photon energy dependence of spectra with the following justifications:

(1) Photon energy dependent stacked Fermi crossings, as seen in SI Fig.  \ref{fig:kz_nodal_hv_dep} provide a rough guide to the inner potential. However, because the observed bands are quasi-2D they do not constrain the inner potential more than $V_0\approx17\pm3eV$. Further attempts were made with other locations in the BZ with similar results. 

\begin{figure*}[h!]
	\centering
	\includegraphics[width=6in]{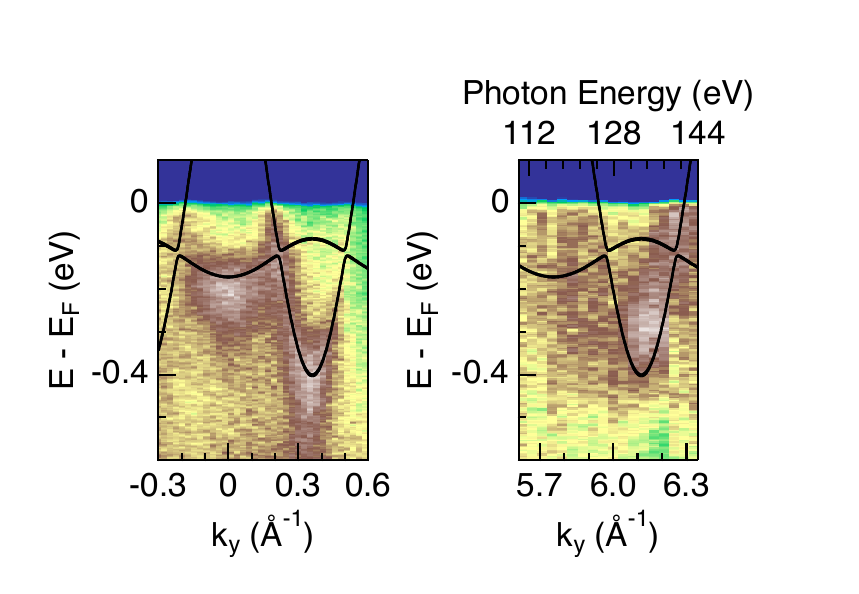}
	\caption{\textbf{Photon energy dependence of $Z-T$ line.} 
 (left) The dispersion measured along the $Z-T$ measured in the thin-edge orientation. (right) Photon energy dependent reconstruction of the $Z-T$ line from data taken in the natural orientation. EDCs were taken at $k_z = \pi/c$ and $k_x=0$ and stacked at many photon energies which reconstructs the $Z-T$ line from measurements on a natural cleave. The perpendicular momentum ($k_y$) is displayed along the bottom axis using an inner potential of $15eV$.
 }
	\label{fig:hv_dep_Z_T}
\end{figure*}

(2) Dispersion tracking as a function of photon energy, as seen in SI Fig.  \ref{fig:hv_dep_Z_T}. A band is reconstructed along $k_{\perp}$ and compared to band structure calculations. $V_0$ is adjusted to yield the best agreement between experiment and theory. 

Strategies (1) and (2) allowed us to constrain the inner potential to be $V_0\approx15\pm1eV$. Any further constraint would require multiple BZs of periodicity to ascertain.

% Fig.  \ref{fig:kz_0_hv_dep} shows the photon energy dependence of photoelectrons integrated near the Fermi level on a naturally cleaved sample (photon energy tunes $k_y$ in this orientation)). Overlaid in black is the DFT calculated Fermi crossings as well as the BZ boundaries, the latter being straight lines. The dashed colored curves represent the momenta probed by a measurement at a given photon energy, shown in the label.

\begin{figure*}[h!]
	\centering
	\includegraphics[width=6in]{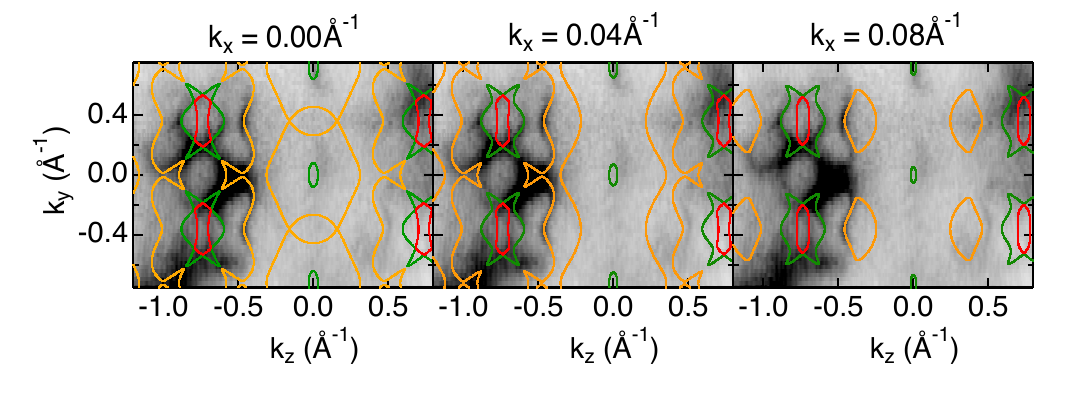}
	\caption{\textbf{Comparison of Thin-edge 144eV Fermi Surface with Nearby Theory} The ARPES data is the same for all panels. It is a constant energy map at the Fermi energy taken on the thin edge with a photon energy of $144eV$. The three panels are overlaid with DFT calculations from different $k_x$ values. }
	\label{fig:Thin_edge_hv_dep}
\end{figure*}

To ascertain which perpendicular momentum is probed with $144eV$ in the thin-edge cleave ($k_x$ in this orientation), we compared the obtained Fermi surface maps with the DFT calculations at multiple values of $k_x$. Three nearby values of $k_x$ are shown in Fig.  \ref{fig:Thin_edge_hv_dep}. Just a $0.08\AA^{-1}$ difference in $k_x$ changes makes an dramatic difference in the Fermiology, and at $k_x=0.08\AA^{-1}$, there are experimental features, particularly along $k_y=0\AA^{-1}$ which are no longer captured by the calculations, as compared to the good fit for $k_x=0\AA^{-1}$ and $k_x=0.04\AA^{-1}$. Because a constant kinetic energy probes a sphere in momentum space, the true difference in measured $k_x$ from the center of the zone to the $k_z$ boundary ($\approx 0.7\AA^{-1}$) is approximately $0.05\AA^{-1}$. That curvature along with the agreement of observed Fermiology to the DFT calculations on the $k_x = 0\AA^{-1}$ plane allow us to determine that a photon energy of $144eV$ probes the $k_x = 0\AA^{-1}$ plane within an uncertainty of $\approx 0.04\AA^{-1}$. %Additionally, the IMFP contribution to uncertainty in $k_\perp$ is $\Delta k_\perp = 0.2\AA^{-1}$ which further enforces that we are at the $k_x =0$ plane as far as the current work can distinguish.

\section{Other ways of visualizing nodal plane}
%Figure \ref{fig:kz_nodal_hv_dep} was created by taking photon energy dependent spectra from $110eV$ to $144eV$. These spectra show evolving Fermiology and matrix elements as a function of photon energy. Near the Gamma plane at $114eV$ we see FS3 extremely bright while FS2 has strongly diminished matrix element. However, at a photon energy of $120eV$, features attributed to FS2 become clearly distinguishable again. With these spectra, the momentum dependent intensity was integrated near the Fermi level and then projected onto $k_y$ vs $k_x$ using equation \ref{eqn:kperp}. 

In the main text, we showed bands merging as individual cuts approach the nodal plane, and in Fig. \ref{fig:kz_nodal_hv_dep} we demonstrate the same for an entire plane. This Figure shows constant energy maps as a function of perpendicular momentum near (top panel) and on (bottom panel) the $k_z$ BZ boundary, along with their respective DFT overlays. This Figure was reconstructed from data collected in the natural cleave geometry.  In this image, the most prominently observed bands are those corresponding to calculated bands FS 4 and 5 (blue, maroon). There is a clear splitting of bands away from the BZ boundary and merging at the BZ, across the whole probed plane.

\begin{figure*}[h!]
	\centering
	\includegraphics[width=4in]{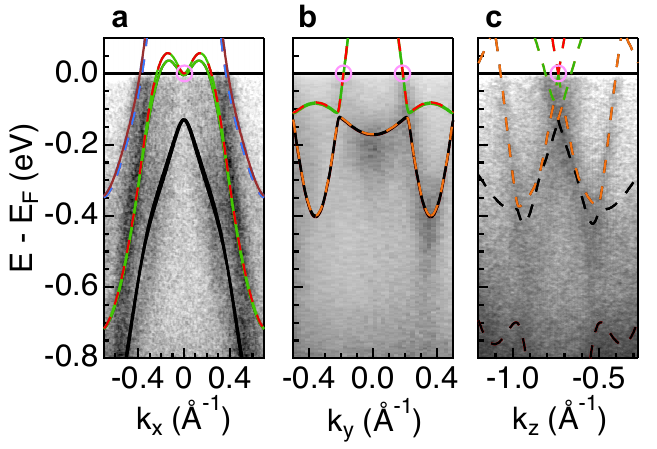}
	\caption{\textbf{Spectra through the Predicted Nodal Points} (a) ARPES spectrum along the $k_x$ direction at a photon energy of 124eV in the natural cleave orientation. (b) ARPES measured along $k_y$ at the zone boundary in the thin-edge orientation. (c) ARPES measured along the $k_z$ axis in the thin-edge orientation. All panels have DFT overlaid colored corresponding to the FS they create. Locations of the predicted nodal points are highlighted with a pink circle.}
	\label{fig:NodalPointsThreeWays}
\end{figure*}

Inspired by Fig. 8 of Y. Quan et al. \cite{Quan2022}, we show the ARPES spectra through the predicted nodal points along $k_x$, $k_y$, and $k_z$ in Fig. \ref{fig:NodalPointsThreeWays}. These spectra are broadly consistent with the predicted electronic structure, which is overlaid and colored corresponding to the Fermi surfaces they create. The predicted SOC protected nodal crossings are highlighted by a pink circle. Fig. \ref{fig:NodalPointsThreeWays} (a) was measured at 124eV in the natural cleave while (b) and (c) were measured in the thin-edge orientation at 144eV. Panel (b) was reconstructed from a deflector scan, as described in the first section of the SI.

\begin{figure*}[h!]
	\centering
	\includegraphics[width=7in]{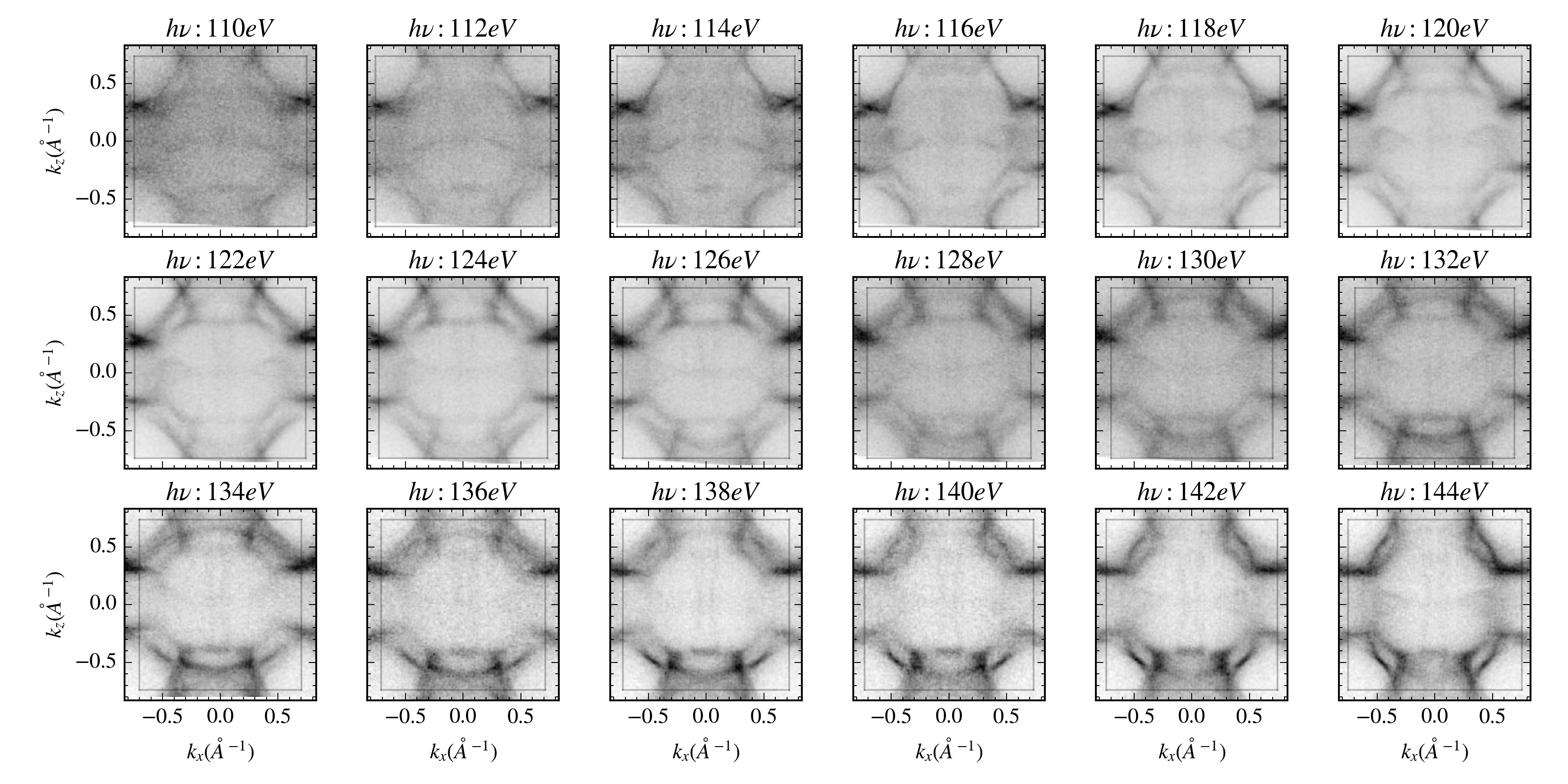}
	\caption{\textbf{Constant energy maps at multiple photon energies} 
 Fermi surface maps integrated $50meV$ around the Fermi energy are shown at a range of photon energies from $110eV$ to $144eV$ in steps of $2eV$. Gray box in each Fig.  outlines the Brillouin zone boundary at that photon energy as determined by an inner potential of $15eV$.
 }
	\label{fig:all_hv_fsmaps}
\end{figure*}

\begin{figure*}[h!]
	\centering
	\includegraphics[width=6in]{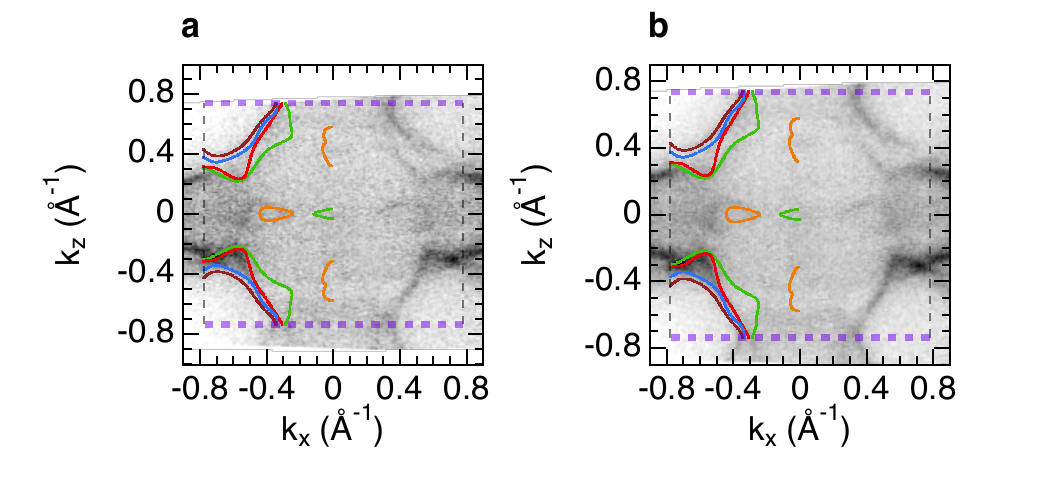}
	\caption{\textbf{Comparison of Energy Integration Windows} 
 (a) Fermi surface map using a photon energy of $116eV$ and integrated $15meV$ around the Fermi level. (b) Same as (a) with an integration window of $50meV$. Both panels have overlaid the calculated Fermi surface on the left half of the data. Black and purple dashed lines indicate the BZ boundary along th $k_x$ and $k_z$ directions respectively.
 }
	\label{fig:integration_test}
\end{figure*}

Fermi surface maps taken between 110 eV and 144 eV are shown in Fig.  \ref{fig:all_hv_fsmaps}.  The nodal plane is oriented horizontally (at constant $k_z$) for each panel, same as similar to Figures in the main text. `X'-shaped features are clearly seen on the nodal plane in most of the Fermi surface maps, as further evidence of the distinct Fermiology of this space group.  These maps were produced using an integration window of $50 meV$ around the Fermi energy, and Fig. \ref{fig:integration_test} shows the Fermi surface map at 116eV produced with different integration windows.  With different integration windows, the consistency/inconsistency with DFT calculations is unchanged, indicating that correspondence with theory is unaffected by this variable.

\section{XPS}

\begin{figure*}[h!]
	\centering
	\includegraphics[width=\linewidth]{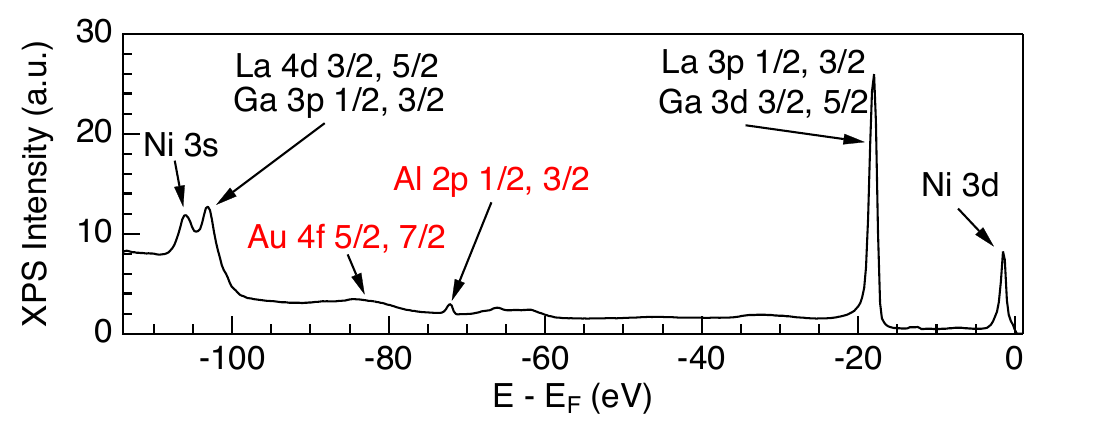}
	\caption{
 \textbf{XPS of LaNiGa$_2$}
 The XPS spectrum was taken at $144eV$. Black annotations show the locations of all known La, Ni, and Ga peaks in the shown binding energy range. The red annotations show the locations of likely contaminant atoms from the growth and sample preparation methods to explain unexpected peaks.
 }
	\label{fig:XPS}
\end{figure*}

In Fig.  \ref{fig:XPS} we show a survey XPS scan of LaNiGa$_2$ taken at a photon energy of $144eV$. This spectrum shows all of the expected peaks for La, Ni, and Ga with the addition of possible contamination of Al from the growth as well as Au which was evaporated on the sample holder prior to cleaving the sample in order to provide a Fermi level reference. 

\end{document}